\documentclass{elsarticle}

\usepackage{amssymb, amsmath}
\usepackage{tabularx, multirow, array}
\usepackage{hyperref}

\usepackage{booktabs}
\usepackage{pgfplotstable}
\pgfplotsset{compat=1.17} 
\usepackage{subfig}

\usepackage{textcomp,color,xcolor,graphicx,colortbl}
\usepackage{ulem}

\newcolumntype{x}{>{\columncolor[gray]{0.8}}l}
\newcolumntype{y}{>{\columncolor[gray]{0.8}}r}
\newcolumntype{z}{>{\columncolor[gray]{0.7}}r}

\journal{Journal of Artificial Intelligence}

\begin{document}

\begin{frontmatter}

\title{Secure Sum Outperforms Homomorphic Encryption in (Current) Collaborative Deep Learning}

\author{Derian Boer\corref{mycorrespondingauthor}
}
\ead{boer@uni-mainz.de}
\author{Stefan Kramer}
\ead{kramer@uni-mainz.de}

\address{Institute of Computer Science, Johannes Gutenberg University Mainz, Staudingerweg 9, 55131 Mainz, Germany}
\cortext[mycorrespondingauthor]{Corresponding author}

\begin{abstract}
Deep learning (DL) approaches are achieving extraordinary results in a wide range of domains, but often require a massive collection of private data. Hence, methods for training neural networks on the joint data of different data owners, that keep each party's input confidential, are called for. We address a specific setting in federated learning, namely that of deep learning from horizontally distributed data with a limited number of parties, where their vulnerable intermediate results have to be processed in a privacy-preserving manner. This setting can be found in medical and healthcare as well as industrial applications.
The predominant scheme for this is based on homomorphic encryption (HE), and it is widely considered to be without alternative. In contrast to this, we demonstrate that a carefully chosen, less complex and computationally less expensive secure sum protocol in conjunction with default secure channels exhibits superior properties in terms of both collusion-resistance and runtime. 
Finally, we discuss several open research questions in the context of collaborative DL, especially regarding privacy risks caused by joint intermediate results.
\end{abstract}

\end{frontmatter}
\clearpage

\section{Introduction} \label{intro}
It is widely acknowledged that the success of deep learning (DL) was made possible, to a great extent, by increasing amounts of collected data.\footnote{Other well-known enabling factors are advances with algorithms, mature software libraries, and the availability of highly parallel hardware.} However, massive data collection is often reserved to the technology giants from the US or China. Therefore, collaborations to train from combined data would be desirable for many companies and institutions.
To complicate matters, many typical applications of deep learning in collaborative settings include sensitive private data that should not be shared in plain text. Collaborating partners can also be competitors, for example, and more or less trusted. Thus, in such cases, ethical and legal privacy concerns have to be considered as constraints. Privacy-preserving data mining and machine learning takes this into account. In this paper, we explicitly focus on a setting with only a relatively small number of parties, as it is prevalent in medical and healthcare as well as industrial applications. In other words, scenarios with a large number of parties, like for instance mobile devices, are beyond the scope of this study.

Privacy-preserving data mining techniques can, by and large, be divided into perturbation and secure multi-party computation (SMC) approaches. Perturbation includes randomization, k-anonymity and l-diversity and downgrading application effectiveness. While these methods target to protect individual's data records from exposure, usually by modifying the input data, and require a trade-off between effectiveness (quality of the output) and privacy, SMC techniques do not affect the effectiveness. 
Instead, special cryptographic communication protocols allow two or more parties to obtain aggregated results from their combined data, such that each of them does not learn any information more than what can be derived from their common output. Yao's protocol solving the Yao's Millionaires' Problem~\cite{Yao82} is seen as the first example of SMC. 
In data mining applications, SMC covers distributed privacy preserving learning which is also known as federated learning and private evaluation. In distributed or collaborative learning, the integrated data are partitioned on multiple parties, which evaluate a function collaboratively and protect their shares. The data may be partitioned horizontally or vertically. 
Federated learning has become its own field in the meantime, with general survey articles \citep{Li20,Yang20,Gadekallu21}, survey articles on specific data types (like natural language \citep{Liu21}), specific learning tasks (like reinforcement learning \citep{Qi21}), and specific properties of data (like being non-IID \cite{Zhu21}). 
Moreover, other encryption schemes such as functional encryption and hybrids with differential privacy \cite{Xu19} are beginning to gain momentum. However, the basic underlying cryptographic question regarding homomorphic encryption vs. secret sharing persists. 
The other field of SMC is private function evaluation, where a client’s input data is processed by a server provided model in such a way that no interpretable data is revealed to any counterparty.

\sloppy{Employing privacy-preserving techniques for stochastic gradient decent (SGD) in deep learning is an intriguing option to develop strong machine learning models with privacy guarantees.} 
To our knowledge, all approaches for deep learning on horizontally partitioned data follow the same procedure: After initialization, all participants train their model locally on one batch each and aggregate their gradients afterwards. The global model parameters are updated with these gradients and then used by every participant in the next iteration. After it has been shown that sharing even small fractions of model parameters in every learning epoch is very vulnerable to generative adversarial networks~\cite{Hit17,Pho18}, present approaches~\cite{Pho18,Li18,Tan19} use a technique to aggregate multiple parties' contributions privately before sharing them. More specifically, they use different types of secure sum protocols. Like many solutions for private evaluation with DNNs, most of the employed secure sum protocols in collaborative DL are based on homomorphic encryption techniques (HE). 
HE enables addition and/or multiplication of encrypted values without a decryption-key and is therefore ubiquitous in privacy preserving algorithms. Their use increases computation and communication costs, but it is feasible in many applications. In collaborative deep learning on several parties' inputs, where much data are processed, a drawback of HE, as used by Phong \textit{et al.}~\cite{Pho18}, is its weakness against collusion attacks when maintaining reasonable costs. Ping Li \textit{et al.}~\cite{Li17,Li18} and Tang \textit{et al.}~\cite{Tan19} provide collusion-resistance by a multi-key and re-encryption scheme, respectively, but require a non-colluding authorization centre (trusted third party). Other collusion-resistant secure aggregation methods as introduced by Chor and Kushilevitz ~\cite{Cho93} seem to be more suitable, because of lower costs and less requirements under two non-restrictive assumptions:
\begin{itemize}
    \item There is no need for a central third-party server, which must not see the global model parameters. 
    \item It is possible to establish pairwise secure channels, e.g., using transport layer security (TLS).
\end{itemize}
The approach we are encouraging can either include a central download server that is given the same security restrictions as any other party and therefore, can see the global parameters after each iteration, or it can schedule the distribution of aggregated results in a decentralized way. We think that this is feasible in practice. A trend of secure sum research is to guarantee security without secure channels thanks to HE~\cite{Jun15,Vu20}. But as it is easy to implement secure paired connections using established TLS connections, the need of being able to communicate without secure channels in complex computations as collaborative DL might be rare. More interestingly, HE could be promising for providing full privacy by working on encrypted intermediate results directly and decrypting the final results only. But to our knowledge, due to insufficient performance, this has not been implemented yet for DNNs and despite missing security guarantees, it has not been proven yet if there is a realistic risk in practice. A recent working paper suggests that a privacy attack on shared intermediate network parameters has little prospect of success, however, it is considered an open problem~\cite{Dou21}.

In this paper, we revive the early but unrecognized approach of N. Schlitter~\cite{Sch08}, which is based on a simple secure sum protocol and does not require any intense public-key encryption. Replacing its secure sum building block by the secure sum algorithm of Urabe \textit{et al.}~\cite{Ura07}, all private inputs are provably indistinguishable from uniformly distributed random numbers even if up to $n-2$ of $n$ parties maliciously collude. There is no need for a trusted third party. Additionally, our experiments and calculations show that this approach combined with basic symmetric key encryption leads to significantly reduced computation costs compared to HE-based approaches in collaborative deep learning. 

The rest of this paper is organized as follows. 
In the next section,
 we provide an overview of
some secure sum protocols with and without HE as a building block of many privacy preserving applications. 
Section~\ref{sec:PPDL} discusses different privacy-preserving deep learning frameworks -- especially on horizontally partitioned data -- and introduces a new advantageous combination of Schlitter's scheme~\cite{Sch08} and the secure sum protocol of Urabe \textit{et al.}~\cite{Ura07}. Additionally, we discuss potential disclosure of intermediate parameters when repeating the protocol. We elaborate the efficiency properties of our suggestion with experimental and theoretical analysis in Section~\ref{sec:impl}.
Finally, Section~\ref{sec:conclusion} concludes the paper and outlines some open research questions. 

\section{Secure Sum Protocols} \label{background}
As a building block for many applications, secure sum is a common example of SMC. In secure sum, $n$ parties have a private input $x_i, i \in \{0,1,...,n-1\}$ each and want to determine the sum $y = \sum_{i=0}^{n-1} x_i$ in such a way that no party learns anything but what can be revealed from its own input and the output $y$.
Presuming no collusion, the problem can be solved straightforwardly as described by Clifton \textit{et al.}~\cite{Cli02}:
\begin{enumerate}
    \item After choosing a limit $l$ known to be greater than $y$, a selected master side draws a random number $r$ from a uniform distribution in the range $[0,l)$.
    \item The master side, denoted as $p_0$, sends $s_0=x_0 + r \mod{l}$ to the second party. 
    \item The parties $p_i \forall i \in \{1,2,...,n-1\}$ sequentially send $s_i = s_{i-1} + x_i \mod{l}$ to party $p_{i+1\mod{n}}$. 
    \item Finally, $p_0$ receives $s_{n-1}$ and broadcasts the result $y=s_{n-1}-r \mod{l}$, which is $r+\sum_{i=0}^{n-1} x_i -r \mod{l} = \sum_{i=0}^{n-1} x_i$, to all other parties.
\end{enumerate}
Clearly, the two adjacent parties $p_{i-1}$ and $p_{i+1\mod{n}}$ of $p_i$ can calculate $x_i$ by subtracting $s_{i-1}$ from $s_i$ if they collude. 
This can be prevented by splitting each $x_i$ into different segments and performing the secure sum protocol for each segment separately in a permuted order.
Chor and Kushilevitz~\cite{Cho93} proposed a $k$-secure protocol on modular addition, which means that no coalition of less than $k<n$ parties can infer any information than what follows from their own input values and the computed sum. It uses $n*k/2$ messages, which is proven to be a tight lower bound for any k-secure oblivious protocol~\cite{Cho93}.

In the following, we briefly outline the secure sum approach of Urabe \textit{et al.}~\cite{Ura07}, which is very similar to the one of Chor and Kushilevitz~\cite{Cho93}.  
It consists of three phases:
\begin{enumerate}
    \item Distribution phase. For each $p_i \in \{1,2,...,n-1\}$ do:
    \begin{itemize}
        \item divide $x_i$ randomly into $n-i$ shares $s_i^j$, such that $\sum\limits_{j=i}^{n-1}s_i^j=x_i$.
        \item send $s_i^j$ to $p_j \forall j\in \{i+1, i+2, ..., n-1\}$. 
    \end{itemize}
    \item Merging phase. For each $p_i \in \{1,2,...,n-1\}$ do:
    \begin{itemize}
        \item calculate $y'_i = \sum\limits_{j=1}^{i}s_j^i$.
        \item send $y'_i$ to $p_0$.
    \end{itemize}
    \item Collection phase. For $p_0$ do:
    \begin{itemize}
        \item calculate the result $y=(x_0+\sum\limits_{i=1}^{n-1} y'_i)$
        \item return / broadcast $y$
    \end{itemize}
\end{enumerate}

\begin{figure}[tb]
	\centering
    \subfloat[distribution phase]{\includegraphics[width=5.5cm]{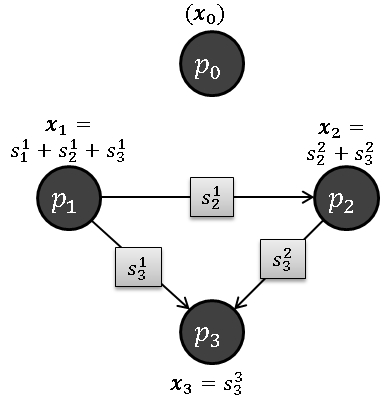}}
    \qquad
    \subfloat[collection phase]{\includegraphics[width=5.5cm]{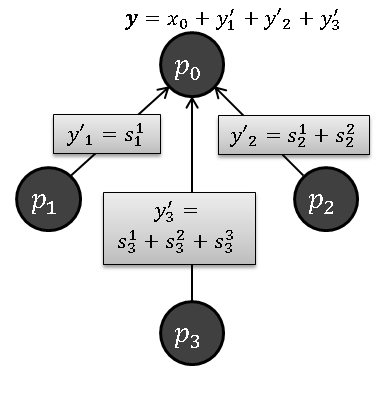}}
    \caption{Communications in secure cum protocol of Urabe \textit{et al.}~\cite{Ura07} for $n=4$}
	\label{fig:ura07}
\end{figure}
Figure \ref{fig:ura07} illustrates the communications for $n=4$. In the end, party $p_0$ receives the final result $y$ and broadcasts it, so that all parties receive it. The communication costs are $n(n-1)b/2$ bits in total for inputs of the size $b$ bits.
By reducing the maximum number of shares of any $x_i$ to $k+1 \leq n-1$ instead of $n-1$, the communication costs can be further reduced to a linear function that depends on a constant $k$. At the same time, the collusion-resistance is reduced to at least $k$ colluding parties. It is sometimes recommended to choose $k$ in the order of $\ln{n}$ to gain a sufficiently small probability of collusion~\cite{Jun15}. 
Jung and Li~\cite{Jun15}
claim that the secure sum protocol of Clifton \textit{et al.}~\cite{Cli02} and other previous protocols require an initialization phase during which participants request keys from key issuers via secure channel, and assume this could be a security risk. 
They introduce a secure product protocol, which is based on the hardness of the discrete logarithm problem, and transfer the private sum problem to a private product problem. In this way, a (computationally) secure sum can be performed without the need for pairwise secure channels.

Other more recently published protocols~\cite{Meh17,Vu20}, fulfilling the property of collusion-resistance for more than two
parties, build on homomorphic encryption techniques (HE). HE allows additions or multiplications of ciphertexts on any two numbers $m_1$ and $m_2$ that have been encrypted by an encryption function $\text{Enc}(x)
$. In additive HE, a practical modulus operator 
 $\oplus$ is known with the property
\begin{equation}
    \text{Enc}(m_1 + m_2) = \text{Enc}(m_1) \oplus \text{Enc}(m_2).
\end{equation}
One example of an additive homomorphic cryptosystem is the assymmetric Paillier cryptosystem~\cite{Pai99}, which is used by Phong \textit{et al.}~\cite{Pho18} in DL. It provides the useful additional property:
\begin{equation}
    \text{Enc}(k*m_1) = k*\text{Enc}(m_1).
\end{equation}
Likewise, in multiplicative HE cryptosystems, there is a modulus operator $\otimes$, such that 
\begin{equation}
    \text{Enc}(m_1 * m_2) = \text{Enc}(m_1) \otimes \text{Enc}(m_2). 
\end{equation}
Mehnaz \textit{et al.}~\cite{Meh17} and Vu \textit{et al.}~\cite{Vu20} employ a variant of the originally multiplicative homomorphic ElGamal Encryption~\cite{Elg85}, that is based on the Diffie-Hellmann encryption~\cite{Dif76} and even fulfills the property $\text{Enc}(m_1 * m_2) = \text{Enc}(m_1) * \text{Enc}(m_2)$.
The former article provides collusion-resistant anonymization, which means that segments are permuted before addition and only $k+1$ parties can determine the origin of segments. The protocol of Vu \textit{et al.}~\cite{Vu20} is based on a multi-party sum function employing a variant of ElGamal encryption, and a Schnorr signature-derived authentication method. In contrast to other approaches, these and some other HE-based protocols do not require pairwise secure channels but can be directly used on public networks.
On the downside, these protocols require many computationally expensive modular exponentiations.
Many other methods, variants of secure aggreation and surveys can be found in the literature, e.g. by Bonawitz \textit{et al.}~\cite{Bon17}, but to our knowledge they have not been applied to DL yet.
We show in section~\ref{sec:impl} that in applications which require many secure aggregations, establishing secure channels can be much cheaper than using HE-based secure sum.

\section{Privacy Preserving Deep Learning}\label{sec:PPDL}
\paragraph{Deep Neural Networks}
We develop a distributed, privacy-preserving learning scheme for deep neural networks. We assume a standard feed-forward neural network trained with back-propagation~\cite{Rum86} and batch normalization, where each update step is performed for a batch $B$ of $|B|$ data records. Let $\Delta W_{ij}^l (d)$ denote the change of the weight $W_{ij}^l$ between the nodes $i$ of layer $l$ and $j$ of layer $l+1$ caused by the data record $d$, the total change of $W_{ij}^l$ in a training step is
\begin{equation}
\label{eq:batch}
    \Delta W_{ij}^l=\sum_{d \in B} \Delta W_{ij}^l (d) = -\eta \sum_{d \in B} G_{ij}^l (d) .
\end{equation}
$\Delta W_{ij}^l (d)$ is a product of a learning rate $\eta$ and the partial derivative $G_{ij}^l (d)$ of the error, e.g., mean squared error, with respect to $W_{ij}^l (d)$.

\paragraph{Privacy Preserving Deep Learning}
Like privacy preserving data mining in general, privacy preserving DL can be divided into different subareas. 
In the field of perturbation methods, Abadi \textit{et al.}~\cite{Aba16}, introduced an ($\epsilon$-)differential private training algorithm for stochastic gradient decent (SGD) that adds noise and accounts for the privacy loss. As typical for perturbation-induced privacy approaches, this leads to a trade-off between accuracy and the individual records' privacy. As we focus on SMC methods in this paper, 
we do not refer to further perturbation-based approaches. 

A lot of articles have been published about private function evaluation with DNNs  \cite{Bac16,Cha17,Li18,Bar06,Ria19,Juv18}. 
A server trains a DNN and offers an evaluation service on a client's data. Neither shall the client receive any more information than the evaluation output nor shall the server be able to infer the client's input data. A common strategy is to approximate all non-linear functions by piece-wise linear ones and employ HE to privately evaluate on the altered DNN. The accuracy loss is low and the computation is feasible in practice, because only two parties and typically just one input vector at once are involved. However, state-of-the-art frameworks rather use hybrids of garbled circuits and secret sharing techniques.

In the field of collaborative DL or federated DL, where the training data are distributed among several data owners, we distinguish between horizontally and vertically partitioned data. 
Chen and Zhong~\cite{Che09} consider vertically partitioned data among two parties, which own different features of common training samples. They use the ElGamal encryption
 scheme to securely compute (1) the piece-wise linear approximation function of the sigmoid activation function of two input summands and (2) the product of two small integers. Both procedures return a random share $s_1$ to one party and another share $s_2$ to the other party, so that $s_1+s_2=y$. These shares and initial weights are propagated through a DNN. At the end of each batch, the random shares are aggregated and the weight updates are shared in plain text and used for the next round. 
Bansal \textit{et al.}~\cite{Ban11} extend this approach by the use of the Paillier encryption-based scalar product algorithm of Wright and Yang~\cite{Wri04}, in such a way that no intermediate results have to be exchanged but random shares. As a result, only the final weights learned by the DNN are leaked in this model. 
The main contribution of this method is its ability to handle arbitrarily partitioned data of two parties. 

\paragraph{Collaborative Deep Learning on Horizontally Partitioned Data}
Recent research in collaborative DL concentrates on DL with horizontally partitioned data. We assume this is due to less complex operation and higher demand in real-world scenarios. 
Many publications (e.g., Shoki \textit{et al.}~\cite{Sho15}) experimentally demonstrate that combining data records of the same kind horizontally can increase the performance of all parties considerably, since it causes more generality and reduces overfitting. 
A training step on a batch of samples consists of a feed-forward and a back-propagation run. The input of a training step are $\eta$, $B$ and the weights from a previous step $\hat{W}$. Let $B_i$ denote the set of samples of $B$ that are owned by party $p_i$ and let $G_i$ be the vector of all partial derivatives (gradients) obtained from $B_i$. To our knowledge, all publications on DL with horizontally partitioned data \cite{Sho15,Sch08,Pho18,Tan19,Meh17,Li17,Li18,Vu20} leverage the property that updates of model parameters are the sum of updates caused by its individual batch samples (see Equation \ref{eq:batch}). Hence, the weight changes caused by individual samples of a party can be calculated independently from the other parties' samples within a batch. The same is done in asynchronous stochastic gradient decent (ASGD) based learning, where the goal is not to preserve privacy but to distribute work load.
The prerequisite is that all parties input the same copy of $\hat{W}$ and preferably the same value of $\eta$ as well. Ideally, $|B_i|=|B|/n$ applies. This leads to the following general steps of a training procedure for collaborative deep learning on horizontally partitioned data:
\begin{enumerate}
    \item Sharing of initial parameters.
    \item\label{start_loop} Local training by each party with input $\hat{W}, B_i$
    \item\label{end_loop} Privacy preserving aggregation of all $G_i$ and update of $\hat{W}$
    \item Repetition of steps \ref{start_loop} to \ref{end_loop} until termination condition is met (see below).
    \end{enumerate}
\begin{figure}
    \centering
    \includegraphics[width=270pt]{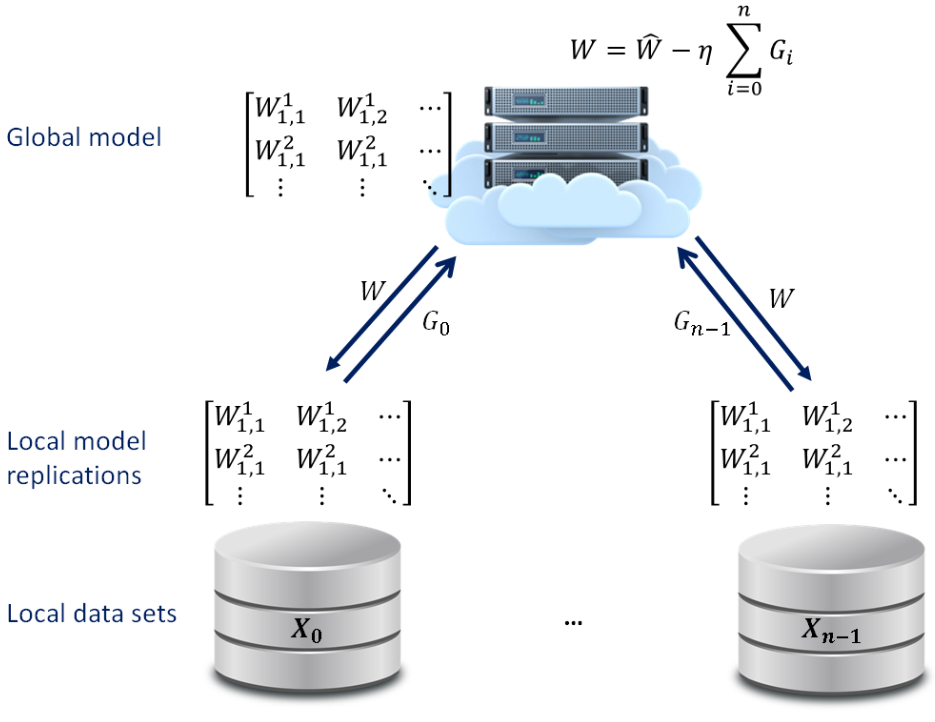}
    \caption{Basic structure of present deep learning on horizontally partitioned data}
    \label{fig:mpdl}
\end{figure}
The initial parameters should be consistent among the parties. Since the parameters are usually independent of the input and randomly generated, any party can broadcast them straightforwardly. 
Subsequently, every party performs a training step on a number of private samples and computes the gradients caused by its private input $B_i$. 
In Step~\ref{end_loop}, a privacy preserving aggregation algorithm is used to securely compute
\begin{equation}
    W = \hat{W}-\eta \sum_{i=0}^{n-1} \sum_{d \in B_i} G_i (d) .
\end{equation}
Figure \ref{fig:mpdl} illustrates the learning loop.
The termination condition can be either a fixed number of iterations or dependent on the loss function. In the latter case, the global loss difference has to be measured regularly. Since the global loss difference is the sum of every party's loss values, it can easily be calculated in a privacy preserving manner by employing a secure sum protocol again.

The main differences of diverse privacy preserving collaborative approaches are how the aggregation of weight updates works. In the following, we briefly point out the differences:

\textbf{Shokri and Shmatikov~\cite{Sho15}} published one of the first discussed articles on collaborative DL with horizontally partitioned training data. They introduce a central semi-honest server that holds the global weights, which the data owners download and update constantly. No cryptographic tool is used for parameter update aggregation, but only a selected fraction of local gradients, e.g., 0.1 or 0.01, after each local batch are asynchronously uploaded. 
These omissions of parameters reduces the accuracy, but the authors claim that the accuracy loss compared to a fully united dataset is acceptable. 

After it has been proven that using this method with setting the fraction of uploaded parameters to even one per cent only is vulnerable to generative adversarial network attacks~\cite{Hit17,Pho18}, \textbf{Phong \textit{et al.}~\cite{Pho18}} applied additive HE (namely Learning With Errors (LWE), and Paillier-based encryption alternatively), before uploading gradients on the server. The server holds the encryted global weights but does not have the private key to decrypt it. Because of the HE property, it can still add encrypted weight updates of the data owners. We point out that, if an honest-but-curious cloud server and any other party collude, the weight updates of any party can be decrypted with the private key known to the other party. Furthermore, the computation is much more expensive because all communications have to be asymmetrically encrypted. 

\textbf{Tang \textit{et al.}~\cite{Tan19}} introduced a homomorphic re-encryption scheme~\cite{Din17} to ASGD-based deep learning, which operates with two cloud servers. One cloud server operates as a data service provider (DSP) and the other one operates as a key transform server (KTS). This protects the weight updates against collusions of any private-key holding learning participant and one of the cloud servers. The authors assume that two cloud servers would not collude for competitive reasons and higher security standards. 
Ping Li \textit{et al.}~\cite{Li17} follow a similar approach.

\sloppy{An earlier, not peer reviewed and not implemented preprint of \textbf{N. Schlitter~\cite{Sch08}} is barely considered in present publications, although it has some beneficial properties.} 
Instead of bothering a third party, the weight updates are added up decentralized using a generic secure sum protocol with an adjustable collusion-resistance parameter $k, 0 < k < n-1$. Replacing the secure sum protocol by the optimized one employed by Urabe \textit{et al.}~\cite{Ura07}, the input of any $p_{a}$ cannot be disclosed if less than all parties but $p_a$ itself collude, which meets the definition of SMC and the same inter-participants collusion-resistance as in the paper of Tang \textit{et al.}~\cite{Tan19}. 
Since the communication is not encrypted, it requires secure channels, e.g., using the Advanced Encryption Standard (AES). In the following, we refer to this composition as \textbf{SUA} (Schlitter-Urabe-AES).

\begin{table}[tb]
    \setlength{\tabcolsep}{3pt}
     \begin{tabular}{ l  c  l}
     \toprule
     Scheme&Parties&collusion-resistance
     \\ \hline
    LWE~\cite{Pho18}&$p_i$'s, DSP&yes, if DSP does not collude with any $p_i$\\ 
    &&and secure channels are used\\
    HRES~\cite{Tan19}&$p_i$'s, DSP, KTS&yes, if DSP and KTS do not collude\\
    SUA&$p_i$'s&yes, if pairwise secure channels are used as intended\\
      \bottomrule
      \end{tabular}
      \caption{Involved parties and security properties of privacy preserving distributed DL frameworks: DSP is the data service provider and KTS the key transform server.}
      \label{tab:collusion-resistance}
\end{table}
Table~\ref{tab:collusion-resistance} contrasts the collusion-resistance of SUA with the HE alternatives of Phong \textit{et al.}~\cite{Pho18} and Tang \textit{et al.}~\cite{Tan19}. In section \ref{sec:impl}, we compare their communication and computation costs.

We note that to our knowledge there is still no approach for more than two parties that matches the strict definition of secure multi computing SMC, that nothing should be disclosed than what can be inferred from the own input and the output, because intermediate results are shared. A summary of an preliminary investigation on the resulting privacy risk is given at the end of Section~\ref{sec:impl}. 
 
\section{Performance Analysis} \label{sec:impl}
In this section, we compare the communication and computation costs of the LWE-based scheme of Phong \textit{et al.}~\cite{Pho18} and the  HRES-based scheme of Tang \textit{et al.}~\cite{Tan19} with the split-and-merge-based scheme based on N. Schliter~\cite{Sch08}. We refer to the scheme based on N. \textbf{S}chliter~\cite{Sch08}
 as SUA, because we modified it in two ways: (1) We replaced the original secure sum protocol by the one of \textbf{U}rabe \textit{et al.}~\cite{Ura07}, because it has lower communication costs and its security properties have been proven. (2) We encrypt all communications in SUA with the Advanced Encryption Standard (\textbf{A}ES), the popular symmetric key scheme that is widely used, e.g., to transfer internet traffic securely. We omit the computationally negligible step of generating and exchanging initial keys and other initialization steps in this paper, since they sum up to far less than a second, e.g., with OpenSSL.

\paragraph{Experimental setup}
As in related articles~\cite{Pho18,Tan19,Sch08}, we consider a default DNN with four layers and 109,386 parameters in total,
represent each parameter by $b=32$ bits and consider a communication bandwidth of 1 Gbps unless otherwise stated. We used the Tensorflow 1.5 library for model training, input images of $28\times 28$ pixels with $|B|=50$, and implemented SUA in Python employing the AES package of the PyCryptodome libary\footnote{\url{https://github.com/Legrandin/pycryptodome}, visited on 18/12/2019} in ECB mode to encrypt all communications. 
We used a recommended~\cite{BSI19,NIST19} key size of 128 bits for AES encryption. 
Unfortunately, the authors of the LWE-based sheme~\cite{Pho18} and HRES-based scheme~\cite{Tan19} did not share their implementations with us, because the methods are patented or code is proprietary, and based on the published information we were not able to reconstruct their reported results.
Thus, we had to compare with the published results and limit our hardware in a way that it cannot be faster than the one used in the original experiments, to achieve a meaningful comparison. 
We ran our experiments on an Intel Xeon W3565 CPU without any hardware accelerations or GPU usage and limited the clock frequency to 3.2 GHz, which is equivalent to the maximum of the CPU used by Tang \textit{et al.}~\cite{Tan19} and slightly lower than the maximum frequency of Phong \textit{et al.}~\cite{Pho18}. We ran all experiments on a single thread unless otherwise stated. However, all compared approaches are similarly parallelizable. 
In their experiment, Tang \textit{et al.}~\cite{Tan19} use a key length of 1024 bits for HRES and a private key with 128 bits. 
For the LWE-based method, the authors~\cite{Pho18} used a similar private key and a public key of 3000 bits.

\paragraph{Communication costs}
\begin{table}[tb]
    \setlength{\tabcolsep}{8.5pt}
     \begin{tabular}{ l  c  c}
     \toprule
     Gradient aggregation and broadcast&communication costs&factor\\ \hline \\[-0.8em]
      LWE~\cite{Pho18}&$\frac{\displaystyle n*n_{\text{lwe}}*\log_2 q}{\displaystyle |W|*b}+\frac{\displaystyle \log_2 q}{\displaystyle b}$&$\approx3.07$\\
    HRES~\cite{Tan19}&$\frac{\displaystyle 2}{\displaystyle |W|}+6$&$\approx6$\\[0.2em]
    SUA&$\frac{\displaystyle n-1}{\displaystyle 4}$&$2.25$\\
      \bottomrule
      \end{tabular}
      \caption{Increase of communication costs in general (second column) and with the given settings (third column)}
      \label{tab:communicationcosts}
\end{table}
For every batch and each party, there is one download and one upload of all gradients.
With the aforementioned parameter settings, the upload size (and download size too) of a non-privacy-preserving distributed learning step is 
\begin{equation}
    |W| * b = 109,386 * 32 \text{ bits} \approx{437.5 \text{ kB}} .
\end{equation}
Table~\ref{tab:communicationcosts} lists the factor of increase of these communication costs of the three privacy-preserving schemes. In the Paillier-based scheme $t$ integers are packed into one plaintext in order to encrypt them more efficiently into one ciphertext, which is encoded with far more than $b$ bits. Therefore, $pad$ padding bits are added to each integer to avoid overflows. As in the original article~\cite{Pho18}, we insert $pad=15$ and the security-sensitive parameter settings $n_{lwe}=3000,q=2^{77}$ into the formulas for an estimation. Since the communication costs of the secure sum by Urabe \textit{et al.}~\cite{Ura07} are $(n(n-1)/2)*b$ bits, it is $((n-1)/2)*b$ bits on average for each party. By passing the distributor role $p_1$ regularly, the communication costs can be balanced uniformly among the participants. Consequently, SUA does not require a central server, because the learning participants receive the sum directly during the aggregation procedure. Hence, aggregation and broadcast are performed in a single step and the factor of increase compared to a corresponding ASGD algorithm is $(n-1)/4$.

\paragraph{Computation costs}
With the assumed network bandwidth of 1 Gbps, the communication runtime for one iteration of a corresponding ASGD algorithm with $n$ distributed cores is:
\begin{equation}
    2 * 437.5\text{ kB} / 1\text{ Gbps} * n = 7.0\text{ ms} * n.
\end{equation}

\begin{table}[tb]
    \setlength{\tabcolsep}{2.4pt}
     \begin{tabular}{l r r r r r| r| r |r}
     \toprule
     Scheme&train.&encr.&add.&decr.&commun.&$\sum$ per $p_i$&DSP+KTS&$\sum$ total\\ \hline
    LWE~\cite{Pho18}&4.6&899.2&-&785.4&214.9&190.4&278.9&2183.0\\
    HRES~\cite{Tan19}&1.3&1112.4&-&584.1&420.0&211.8&3496.0&5613.8\\
    SUA&1.5&118.2&69.9&108.6&157.5&25.1&-&455.7\\
    Non priv.&1.5&-&-&-&70.0&7.2&1.0&72.5\\
      \bottomrule
      \end{tabular}
      \caption{Runtime (ms) for one iteration ($W=109,386$) with $n=10$, 1 thread}
      \label{tab:computationcosts}
\end{table}
The communication costs of the privacy-preserving increase by the factors in Table~\ref{tab:communicationcosts}.

Table~\ref{tab:computationcosts} contrasts the total runtimes of the LWE-based scheme by Phong \textit{et al.}~\cite{Pho18}, the HRES-based scheme by Tang \textit{et al.}~\cite{Tan19} and our measurements for SUA and a non-private ASGD iteration for $n=10$. The sum per learning participants ($\sum$ per $p_i$) consists of the local training, encryption and decryption, and communication for one $p_i$ taking one thread per learning participant. In case of SUA, the costs for the secure sum protocol count in this measure, because the addition is performed decentralized. Whereas in the other cases we exclude it, because a DSP does the addition, which has usually more computational capacity according to the argumentation of Tang \textit{et al.}~\cite{Tan19}. For the same reason, we also exclude the operations performed by the KTS in HRES here. 
In our experiment in Table \ref{tab:computationcosts}, SUA's cryptographic overhead leads to a 6-fold increase of computation costs compared to a non-private federated learning solution. But even when considering the communication and participants’ computation only, SUA causes 4-5 times less runtime compared to both HE-based schemes for $n=10$.

\begin{figure}[tb]
\caption{Total runtime (without operations done by DSP and KTS) for one iteration ($|W|=109,386$) using \textit{1 thread} in total}
\label{fig:runtimes}
\begin{tikzpicture}
		\begin{axis}[width=0.75\textwidth,height=0.29\textheight,xlabel={$n$},ylabel={runtime in s},legend pos=outer north east] 
		    legend style={at={(0.03,0.5)},anchor=east};
			\addplot table[x=n,y=LWE, col sep=comma]{runtimes.csv};
            \addlegendentry{LWE~\cite{Pho18}}
            \addplot table[x=n,y=HRES, col sep=comma]{runtimes.csv};
            \addlegendentry{HRES~\cite{Tan19}}
            \addplot table[x=n,y=SchUraAES, col sep=comma]{runtimes.csv};
            \addlegendentry{SUA ($k=n-1$)}
            \addplot table[x=n,y=SUAkn2, col sep=comma]{runtimes.csv};
            \addlegendentry{SUA ($k=\lceil n/2 \rceil $)}
            \addplot table[x=n,y=Non-Private, col sep=comma]{runtimes.csv};
            \addlegendentry{Non private}
		\end{axis}
\end{tikzpicture}
\end{figure}
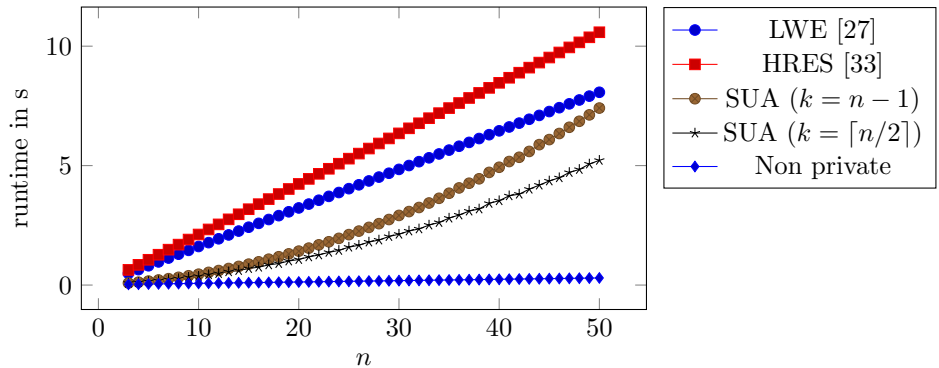
Figure~\ref{fig:runtimes} contrasts the total runtime calculated for one iteration for different values of $n$ of the centralized HE-based schemes, ignoring costs of DSP and KTS, and two variants of SUA: with $k=n-1$ for maximum collusion security and with $k=\lceil n/2\rceil$ additionally, which preserves privacy even if any half of the participants collude. We calculated the runtimes for encryption, decryption and communication in the cloud server approaches by assuming constant time per $p_i$ and assumed a constant local time for training because $|B|$ remains constant. As expected, the runtime of SUA per $p_i$ increases with growing $n$ because the costs for encrypted communication per $p_i$ are proportional to $n-1$. However, the level of collusion resistance can be limited to a fixed security parameter $k$, such that the protocol is secure against any collusion of up to $k$ instead of $n-1$ participants. With a fixed $k$, runtime for the secure sum protocol and its communications per $p_i$ remain constant, making SUA very scalable if dropouts and collusions of a larger scale are very unlikely. For $n \leq 66, k=n-1$ and $n \leq 100,k=\lceil n/2 \rceil$, SUA remains the faster privacy-preserving procedure, even when ignoring costs for the DSP and KTS.

\paragraph{Network latency and bandwidth}
\begin{table}[tb]
\centering
    \setlength{\tabcolsep}{2.4pt}
     \begin{tabular}{x| r r >{\textit}r r y| z}
     \toprule
     Bandwidth(Mbits/s)&train.&encr.&add.&decr.&commun.&$\sum$ total\\ \hline
    125&1.5&118.2&69.9&108.6&1260.0&1567.3\\
    250&1.5&118.2&69.9&108.6&630.3&927.3\\
    500&1.5&118.2&69.9&108.6&315.0&612.3\\
    1000&1.5&118.2&69.9&108.6&157.5&454.8\\
    2000&1.5&118.2&69.9&108.6&78.8&376.1\\
    4000&1.5&118.2&69.9&108.6&39.4&336.7\\
      \bottomrule
      \end{tabular}
      \caption{Runtime (ms) of one iteration ($W=109,386$) of SUA with different bandwidths and $n=10$, 1 thread}
      \label{tab:bandwidth}
\end{table}
In the previous paragraphs, we assumed a bandwidth of 1 Gbps for comparison to the work of Phong \textit{et al.}~\cite{Pho18} and Tang \textit{et al.}~\cite{Tan19}. However, very different network conditions are conceivable.
Note that the lowest bandwidth in a network determines the runtime among all participants, because broadcasting of the next iteration's parameter depends on the processing of all inputs.
Table~\ref{tab:bandwidth} displays the communication costs for different bandwidths in relation to other computation costs of SUA. Although it can been seen that a low bandwidth decreases the overall performance substantially (because the communication costs increase anti-proportionally), a varying moderate bandwidth is not a crucial factor in comparison to the others, because it affects all methods similarly. 

In the previous paragraphs, we omitted delays in communication caused by network and hardware constraints for simplicity -- as in the work of Phong \textit{et al.}~\cite{Pho18} and Tang \textit{et al.}~\cite{Tan19}. A measure for such delays, e.g., round trip delay, is latency: the time it takes for a minimal amount of data to get to its destination across the network (and back again).
Many factors influence the latency: physical distance between servers and computers, propagation delay, internet connection type, router properties, etc. We find distance to be the dominating factor here and, hence, assume the delay of parallel communications not to be cumulative: Let $d$ be the maximum delay of arrival of a message sent from any party to any other one. In a non-private setting, the total delay of one iteration is $2d$, since $n$ uploads and subsequently $n$ downloads are performed in parallel. The same applies to the LWE scheme. For HRES it is $2d+2d'$, where $d‘$ is the delay from DSP to KTS and vice versa. For SUA, it is $3d$ because one iteration of the protocol includes three phases of parallel communications (distribution, collection, broadcast).

\begin{table}[tb]
\centering
    \setlength{\tabcolsep}{2.4pt}
     \begin{tabular}{x c c r r r| y| z}
     \toprule
     $d$&train.&encr.&add.&decr.&commun.&delay ($=3d$)&$\sum$ total\\ \hline
    10&1.5&118.2&69.9&108.6&157.5&30&484.8\\
    20&1.5&118.2&69.9&108.6&157.5&60&514.8\\
    40&1.5&118.2&69.9&108.6&157.5&120&574.8\\
    80&1.5&118.2&69.9&108.6&157.5&240&694.8\\
    160&1.5&118.2&69.9&108.6&157.5&480&934.8\\
      \bottomrule
      \end{tabular}
      \caption{Runtime (ms) of one iteration ($W=109,386$) of SUA with different latencies and $n=10$, 1 thread}
      \label{tab:delay}
\end{table}
Table~\ref{tab:delay} shows how different possible values for $d$ influence the total performance, taking SUA as an example. Like a low bandwidth, a high latency has a substantial, almost linear impact on the overall runtime, but a moderate latency is a negligible factor when it comes to comparison of the three protocols. In practice, we measured a TCP/IP round trip delay ("ping") of 16ms from Mainz, Germany, to Oxford\footnote{https://www.ox.ac.uk}, UK (about 700km  distance), e.g.. If a connection is checked every time before data are sent, $d$ may be estimated by $1.5$ times the round trip delay.

\section{Potential Disclosure of Intermediate Parameters When Repeating the Protocol}\label{intermediate}
While our employed aggregation protocol is proven to be secure in a single step, the risk of violating privacy by sharing aggregated intermediate results, i.e., repeating the protocol over a potentially large number of steps, has not been assessed yet. To investigate the level of security for subsequent iterations, one starting point is to look into the system of equations that needs to be solved by an adversary to recover any party's dataset. The first step is to focus on the exact version of the problem, where the adversary is interested in determining the exact dataset, rather than gaining partial knowledge about it. A working paper by Dousti~\cite{Dou21} sheds some light on the issue. A way of forming such a system of equations is given there, and it is demonstrated that it is highly non-linear and even for the simplest possible problem instance not solvable by common state-of-the-art solver implementations such as SciPy and SageMath. In general, as shown by Blum \textit{et al.}~\cite{Blu89}, solving a non-linear system (even if it is only quadratic) is NP-complete in the worst case. 
From the point of view of numerical mathematics\footnote{Personal communication with Markus Bachmayr (2021).}, there are no general known statements about systems of equations of this kind in the literature. Depending on the chosen initial values and approach, there is always the risk of running into problems with convergence or overflows. 
In general, the problem (clearly) is an inverse problem, where the existence, uniqueness and stability of the reconstruction are not evident. Therefore, it might make sense not to demand exact identities, but reformulate it as a least squares minimization problem. With this, one can usually expect better robustness. However, the issues with having to solve a highly nonlinear equation system remain.
Finally, the study of Dousti~\cite{Dou21} describes that it is probably infeasible to even form the system if the underlying ANN uses a randomized learning approach such as stochastic gradient decent (SGD), because it is again highly sensitive to initial conditions. All of this indicates that a solution to this problem may be infeasible in practice, however, no proof can be given at this point.

Overall, if the risk of reconstructing datasets from several intermediate parameter should turn out to be high, future work could exploit the HE property to prevent disclosure of \textit{any} intermediate results. HE has, in principle, the opportunity to perform the whole training on encrypted data. 
Therefore, more efficient approaches for this needed to be developed. HE techniques for symmetric keys might be an option. Another open question concerns vertically or arbitrarily partitioned data for more than two parties.

\section{Conclusion}\label{sec:conclusion}
Homomorphic encryption is a powerful, versatile tool, but applied to asymmetric encryption, it is often computationally costly. In the field of SMC, other techniques should not be neglected. We demonstrate the advantages of alternatives in the example of DL on horizontally partitioned data.
Instead of 2.5 hours with 20 threads (10 cores)~\cite{Pho18} with HE based protocols, we can train a default DNN in 6 to 35 minutes (for $n \in \{3,4,...,20\}$) with 4 threads employing a simple but even fully collusion secure sum protocol. Since we target industrial and institutional B2B settings, where dropouts of participants are very rare, future work should take further secure aggregation algorithms into account that are suitable for peer-to-peer environments.

The risk of violating privacy by sharing aggregated \textit{intermediate results} has not conclusively been evaluated yet. HE has the chance to perform the whole training on encrypted data. Therefore, more efficient approaches should be developed. 
On the other hand, there is evidence that an abuse of intermediate parameters in order to exactly reconstruct inputs is infeasible in practice, especially with SGD, because of the non-linearity of such attacks. More research is needed to prove or disprove this. Another unsolved problem concerns vertically or arbitrarily partitioned data for more than two parties.

\section*{Acknowledgements}
This research did not receive any specific grant from funding agencies in the public, commercial, or not-for-profit sectors.

\bibliography{references}

\begin{thebibliography}{40}
\expandafter\ifx\csname natexlab\endcsname\relax\def\natexlab#1{#1}\fi
\providecommand{\url}[1]{\texttt{#1}}
\providecommand{\href}[2]{#2}
\providecommand{\path}[1]{#1}
\providecommand{\DOIprefix}{doi:}
\providecommand{\ArXivprefix}{arXiv:}
\providecommand{\URLprefix}{URL: }
\providecommand{\Pubmedprefix}{pmid:}
\providecommand{\doi}[1]{\href{http://dx.doi.org/#1}{\path{#1}}}
\providecommand{\Pubmed}[1]{\href{pmid:#1}{\path{#1}}}
\providecommand{\bibinfo}[2]{#2}
\ifx\xfnm\relax \def\xfnm[#1]{\unskip,\space#1}\fi
\bibitem[{Abadi et~al.(2016)Abadi, Chu, Goodfellow, McMahan, Mironov, Talwar \&
  Zhang}]{Aba16}
\bibinfo{author}{Abadi, M.}, \bibinfo{author}{Chu, A.},
  \bibinfo{author}{Goodfellow, I.~J.}, \bibinfo{author}{McMahan, H.~B.},
  \bibinfo{author}{Mironov, I.}, \bibinfo{author}{Talwar, K.}, \&
  \bibinfo{author}{Zhang, L.} (\bibinfo{year}{2016}).
\newblock \bibinfo{title}{Deep learning with differential privacy}.
\newblock In \bibinfo{editor}{E.~R. Weippl},
  \bibinfo{editor}{S.~Katzenbeisser}, \bibinfo{editor}{C.~Kruegel},
  \bibinfo{editor}{A.~C. Myers}, \& \bibinfo{editor}{S.~Halevi} (Eds.), {\it
  \bibinfo{booktitle}{Proceedings of the 2016 {ACM} {SIGSAC} Conference on
  Computer and Communications Security, Vienna, Austria, October 24-28,
  2016}\/} (pp. \bibinfo{pages}{308--318}).
\newblock \bibinfo{publisher}{{ACM}}.
\newblock \DOIprefix\doi{10.1145/2976749.2978318}.
\bibitem[{Bansal et~al.(2011)Bansal, Chen \& Zhong}]{Ban11}
\bibinfo{author}{Bansal, A.}, \bibinfo{author}{Chen, T.}, \&
  \bibinfo{author}{Zhong, S.} (\bibinfo{year}{2011}).
\newblock \bibinfo{title}{Privacy preserving back-propagation neural network
  learning over arbitrarily partitioned data}.
\newblock {\it \bibinfo{journal}{Neural Computing and Applications}\/},  {\it
  \bibinfo{volume}{20}\/}, \bibinfo{pages}{143--150}.
  \DOIprefix\doi{10.1007/s00521-010-0346-z}.
\bibitem[{Barker \& Roginsky(2019)}]{NIST19}
\bibinfo{author}{Barker, E.~B.}, \& \bibinfo{author}{Roginsky, A.~L.}
  (\bibinfo{year}{2019}).
\newblock {\it \bibinfo{title}{Transitioning the Use of Cryptographic
  Algorithms and Key Lengths}\/}.
\newblock \bibinfo{type}{Technical Report} \bibinfo{number}{800-131A Rev. 2} US
  National Institute of Standards and Technology (NIST).
\newblock \DOIprefix\doi{10.6028/NIST.SP.800-131Ar2}.
\bibitem[{Barni et~al.(2006)Barni, Orlandi \& Piva}]{Bar06}
\bibinfo{author}{Barni, M.}, \bibinfo{author}{Orlandi, C.}, \&
  \bibinfo{author}{Piva, A.} (\bibinfo{year}{2006}).
\newblock \bibinfo{title}{A privacy-preserving protocol for
  neural-network-based computation}.
\newblock In \bibinfo{editor}{S.~Voloshynovskiy},
  \bibinfo{editor}{J.~Dittmann}, \& \bibinfo{editor}{J.~J. Fridrich} (Eds.),
  {\it \bibinfo{booktitle}{Proceedings of the 8th workshop on Multimedia {\&}
  Security, MM{\&}Sec 2006, Geneva, Switzerland, September 26-27, 2006}\/} (pp.
  \bibinfo{pages}{146--151}).
\newblock \bibinfo{publisher}{{ACM}}.
\newblock \DOIprefix\doi{10.1145/1161366.1161393}.
\bibitem[{Blum et~al.(1989)Blum, Shub \& Smale}]{Blu89}
\bibinfo{author}{Blum, L.}, \bibinfo{author}{Shub, M.}, \&
  \bibinfo{author}{Smale, S.} (\bibinfo{year}{1989}).
\newblock \bibinfo{title}{{On a theory of computation and complexity over the
  real numbers: $NP$- completeness, recursive functions and universal
  machines}}.
\newblock {\it \bibinfo{journal}{Bulletin (New Series) of the American
  Mathematical Society}\/},  {\it \bibinfo{volume}{21}\/}, \bibinfo{pages}{1 --
  46}. \URLprefix \url{https://doi.org/}. \DOIprefix\doi{bams/1183555121}.
\bibitem[{Bonawitz et~al.(2017)Bonawitz, Ivanov, Marcedone, Kreuter, McMahan,
  Patel, Ramage, Segal \& Seth}]{Bon17}
\bibinfo{author}{Bonawitz, K.~A.}, \bibinfo{author}{Ivanov, V.},
  \bibinfo{author}{Marcedone, A.}, \bibinfo{author}{Kreuter, B.},
  \bibinfo{author}{McMahan, H.~B.}, \bibinfo{author}{Patel, S.},
  \bibinfo{author}{Ramage, D.}, \bibinfo{author}{Segal, A.}, \&
  \bibinfo{author}{Seth, K.} (\bibinfo{year}{2017}).
\newblock \bibinfo{title}{Practical secure aggregation for privacy-preserving
  machine learning}.
\newblock In {\it \bibinfo{booktitle}{CCS}\/}.
\newblock \URLprefix \url{https://eprint.iacr.org/2017/281.pdf}.
\bibitem[{Chabanne et~al.(2017)Chabanne, de~Wargny, Milgram, Morel \&
  Prouff}]{Cha17}
\bibinfo{author}{Chabanne, H.}, \bibinfo{author}{de~Wargny, A.},
  \bibinfo{author}{Milgram, J.}, \bibinfo{author}{Morel, C.}, \&
  \bibinfo{author}{Prouff, E.} (\bibinfo{year}{2017}).
\newblock \bibinfo{title}{Privacy-preserving classification on deep neural
  network}.
\newblock {\it \bibinfo{journal}{{IACR} Cryptology ePrint Archive}\/},  {\it
  \bibinfo{volume}{2017}\/}, \bibinfo{pages}{35}.
\bibitem[{Chen \& Zhong(2009)}]{Che09}
\bibinfo{author}{Chen, T.}, \& \bibinfo{author}{Zhong, S.}
  (\bibinfo{year}{2009}).
\newblock \bibinfo{title}{Privacy-preserving backpropagation neural network
  learning}.
\newblock {\it \bibinfo{journal}{{IEEE} Trans. Neural Networks}\/},  {\it
  \bibinfo{volume}{20}\/}, \bibinfo{pages}{1554--1564}.
  \DOIprefix\doi{10.1109/TNN.2009.2026902}.
\bibitem[{Chor \& Kushilevitz(1993)}]{Cho93}
\bibinfo{author}{Chor, B.}, \& \bibinfo{author}{Kushilevitz, E.}
  (\bibinfo{year}{1993}).
\newblock \bibinfo{title}{A communication-privacy tradeoff for modular
  addition}.
\newblock {\it \bibinfo{journal}{Information Processing Letters}\/},  {\it
  \bibinfo{volume}{45}\/}, \bibinfo{pages}{205--210}. \URLprefix
  \url{https://www.sciencedirect.com/science/article/pii/002001909390120X}.
  \DOIprefix\doi{https://doi.org/10.1016/0020-0190(93)90120-X}.
\bibitem[{Clifton et~al.(2002)Clifton, Kantarcioglu, Vaidya, Lin \&
  Zhu}]{Cli02}
\bibinfo{author}{Clifton, C.}, \bibinfo{author}{Kantarcioglu, M.},
  \bibinfo{author}{Vaidya, J.}, \bibinfo{author}{Lin, X.}, \&
  \bibinfo{author}{Zhu, M.~Y.} (\bibinfo{year}{2002}).
\newblock \bibinfo{title}{Tools for privacy preserving distributed data
  mining}.
\newblock {\it \bibinfo{journal}{SIGKDD Explor. Newsl.}\/},  {\it
  \bibinfo{volume}{4}\/}, \bibinfo{pages}{28–34}.
  \DOIprefix\doi{10.1145/772862.772867}.
\bibitem[{Diffie \& Hellman(1976)}]{Dif76}
\bibinfo{author}{Diffie, W.}, \& \bibinfo{author}{Hellman, M.~E.}
  (\bibinfo{year}{1976}).
\newblock \bibinfo{title}{New directions in cryptography}.
\newblock {\it \bibinfo{journal}{{IEEE} Trans. Information Theory}\/},  {\it
  \bibinfo{volume}{22}\/}, \bibinfo{pages}{644--654}.
  \DOIprefix\doi{10.1109/TIT.1976.1055638}.
\bibitem[{Ding et~al.(2017)Ding, Yan \& Deng}]{Din17}
\bibinfo{author}{Ding, W.}, \bibinfo{author}{Yan, Z.}, \&
  \bibinfo{author}{Deng, R.~H.} (\bibinfo{year}{2017}).
\newblock \bibinfo{title}{Encrypted data processing with homomorphic
  re-encryption}.
\newblock {\it \bibinfo{journal}{Information Sciences}\/},  {\it
  \bibinfo{volume}{409}\/}, \bibinfo{pages}{35--55}.
  \DOIprefix\doi{10.1016/j.ins.2017.05.004}.
\bibitem[{Dousti(2021)}]{Dou21}
\bibinfo{author}{Dousti, M.~S.} (\bibinfo{year}{2021}).
\newblock \bibinfo{title}{Neural networks, inside out: Solving for inputs given
  parameters (a preliminary investigation)}.
\newblock \href{http://arxiv.org/abs/2110.03649}{\tt arXiv:2110.03649}.
\bibitem[{Gadekallu et~al.(2021)Gadekallu, Pham, Huynh-The, Bhattacharya,
  Maddikunta \& Liyanage}]{Gadekallu21}
\bibinfo{author}{Gadekallu, T.~R.}, \bibinfo{author}{Pham, Q.-V.},
  \bibinfo{author}{Huynh-The, T.}, \bibinfo{author}{Bhattacharya, S.},
  \bibinfo{author}{Maddikunta, P. K.~R.}, \& \bibinfo{author}{Liyanage, M.}
  (\bibinfo{year}{2021}).
\newblock \bibinfo{title}{Federated learning for big data: A survey on
  opportunities, applications, and future directions}.
\newblock \href{http://arxiv.org/abs/2110.04160}{\tt arXiv:2110.04160}.
\bibitem[{Gamal(1985)}]{Elg85}
\bibinfo{author}{Gamal, T.~E.} (\bibinfo{year}{1985}).
\newblock \bibinfo{title}{A public key cryptosystem and a signature scheme
  based on discrete logarithms}.
\newblock {\it \bibinfo{journal}{{IEEE} Trans. Information Theory}\/},  {\it
  \bibinfo{volume}{31}\/}, \bibinfo{pages}{469--472}.
  \DOIprefix\doi{10.1109/TIT.1985.1057074}.
\bibitem[{{German Federal Office for Information Security (BSI)}(2019)}]{BSI19}
\bibinfo{author}{{German Federal Office for Information Security (BSI)}}
  (\bibinfo{year}{2019}).
\newblock {\it \bibinfo{title}{Cryptographic Mechanisms: Recommendations and
  Key Lengths}\/}.
\newblock \bibinfo{type}{Technical Report}.
\newblock \bibinfo{note}{Version 2019-01, retrieved from
  \url{https://www.bsi.bund.de/SharedDocs/Downloads/EN/BSI/Publications/TechGuidelines/TG02102/BSI-TR-02102-1.html}}.
\bibitem[{Gilad{-}Bachrach et~al.(2016)Gilad{-}Bachrach, Dowlin, Laine, Lauter,
  Naehrig \& Wernsing}]{Bac16}
\bibinfo{author}{Gilad{-}Bachrach, R.}, \bibinfo{author}{Dowlin, N.},
  \bibinfo{author}{Laine, K.}, \bibinfo{author}{Lauter, K.~E.},
  \bibinfo{author}{Naehrig, M.}, \& \bibinfo{author}{Wernsing, J.}
  (\bibinfo{year}{2016}).
\newblock \bibinfo{title}{Cryptonets: Applying neural networks to encrypted
  data with high throughput and accuracy}.
\newblock In \bibinfo{editor}{M.~Balcan}, \& \bibinfo{editor}{K.~Q. Weinberger}
  (Eds.), {\it \bibinfo{booktitle}{Proceedings of the 33nd International
  Conference on Machine Learning, {ICML} 2016, New York City, NY, USA, June
  19-24, 2016}\/} (pp. \bibinfo{pages}{201--210}).
\newblock \bibinfo{publisher}{JMLR.org} volume~\bibinfo{volume}{48} of {\it
  \bibinfo{series}{{JMLR} Workshop and Conference Proceedings}\/}.
\bibitem[{Hitaj et~al.(2017)Hitaj, Ateniese \& P{\'{e}}rez{-}Cruz}]{Hit17}
\bibinfo{author}{Hitaj, B.}, \bibinfo{author}{Ateniese, G.}, \&
  \bibinfo{author}{P{\'{e}}rez{-}Cruz, F.} (\bibinfo{year}{2017}).
\newblock \bibinfo{title}{Deep models under the {GAN:} information leakage from
  collaborative deep learning}.
\newblock In \bibinfo{editor}{B.~M. Thuraisingham}, \bibinfo{editor}{D.~Evans},
  \bibinfo{editor}{T.~Malkin}, \& \bibinfo{editor}{D.~Xu} (Eds.), {\it
  \bibinfo{booktitle}{Proceedings of the 2017 {ACM} {SIGSAC} Conference on
  Computer and Communications Security, {CCS} 2017, Dallas, TX, USA, October 30
  - November 03, 2017}\/} (pp. \bibinfo{pages}{603--618}).
\newblock \bibinfo{publisher}{{ACM}}.
\newblock \DOIprefix\doi{10.1145/3133956.3134012}.
\bibitem[{Jung et~al.(2015)Jung, Li \& Wan}]{Jun15}
\bibinfo{author}{Jung, T.}, \bibinfo{author}{Li, X.}, \& \bibinfo{author}{Wan,
  M.} (\bibinfo{year}{2015}).
\newblock \bibinfo{title}{Collusion-tolerable privacy-preserving sum and
  product calculation without secure channel}.
\newblock {\it \bibinfo{journal}{{IEEE} Trans. Dependable Sec. Comput.}\/},
  {\it \bibinfo{volume}{12}\/}, \bibinfo{pages}{45--57}.
  \DOIprefix\doi{10.1109/TDSC.2014.2309134}.
\bibitem[{Juvekar et~al.(2018)Juvekar, Vaikuntanathan \& Chandrakasan}]{Juv18}
\bibinfo{author}{Juvekar, C.}, \bibinfo{author}{Vaikuntanathan, V.}, \&
  \bibinfo{author}{Chandrakasan, A.} (\bibinfo{year}{2018}).
\newblock \bibinfo{title}{{GAZELLE:} {A} low latency framework for secure
  neural network inference}.
\newblock In \bibinfo{editor}{W.~Enck}, \& \bibinfo{editor}{A.~P. Felt} (Eds.),
  {\it \bibinfo{booktitle}{27th {USENIX} Security Symposium, {USENIX} Security
  2018, Baltimore, MD, USA, August 15-17, 2018}\/} (pp.
  \bibinfo{pages}{1651--1669}).
\newblock \bibinfo{publisher}{{USENIX} Association}.
\bibitem[{Li et~al.(2018)Li, Li, Huang, Gao, Chen \& Chen}]{Li18}
\bibinfo{author}{Li, P.}, \bibinfo{author}{Li, J.}, \bibinfo{author}{Huang,
  Z.}, \bibinfo{author}{Gao, C.}, \bibinfo{author}{Chen, W.}, \&
  \bibinfo{author}{Chen, K.} (\bibinfo{year}{2018}).
\newblock \bibinfo{title}{Privacy-preserving outsourced classification in cloud
  computing}.
\newblock {\it \bibinfo{journal}{Cluster Computing}\/},  {\it
  \bibinfo{volume}{21}\/}, \bibinfo{pages}{277--286}.
  \DOIprefix\doi{10.1007/s10586-017-0849-9}.
\bibitem[{Li et~al.(2017)Li, Li, Huang, Li, Gao, Yiu \& Chen}]{Li17}
\bibinfo{author}{Li, P.}, \bibinfo{author}{Li, J.}, \bibinfo{author}{Huang,
  Z.}, \bibinfo{author}{Li, T.}, \bibinfo{author}{Gao, C.},
  \bibinfo{author}{Yiu, S.}, \& \bibinfo{author}{Chen, K.}
  (\bibinfo{year}{2017}).
\newblock \bibinfo{title}{Multi-key privacy-preserving deep learning in cloud
  computing}.
\newblock {\it \bibinfo{journal}{Future Generation Comp. Syst.}\/},  {\it
  \bibinfo{volume}{74}\/}, \bibinfo{pages}{76--85}.
  \DOIprefix\doi{10.1016/j.future.2017.02.006}.
\bibitem[{Li et~al.(2020)Li, Sahu, Talwalkar \& Smith}]{Li20}
\bibinfo{author}{Li, T.}, \bibinfo{author}{Sahu, A.~K.},
  \bibinfo{author}{Talwalkar, A.}, \& \bibinfo{author}{Smith, V.}
  (\bibinfo{year}{2020}).
\newblock \bibinfo{title}{Federated learning: Challenges, methods, and future
  directions}.
\newblock {\it \bibinfo{journal}{{IEEE} Signal Process. Mag.}\/},  {\it
  \bibinfo{volume}{37}\/}, \bibinfo{pages}{50--60}. \URLprefix
  \url{https://doi.org/10.1109/MSP.2020.2975749}.
  \DOIprefix\doi{10.1109/MSP.2020.2975749}.
\bibitem[{Liu et~al.(2021)Liu, Ho, Wang, Gao, Jin \& Zhang}]{Liu21}
\bibinfo{author}{Liu, M.}, \bibinfo{author}{Ho, S.}, \bibinfo{author}{Wang,
  M.}, \bibinfo{author}{Gao, L.}, \bibinfo{author}{Jin, Y.}, \&
  \bibinfo{author}{Zhang, H.} (\bibinfo{year}{2021}).
\newblock \bibinfo{title}{Federated learning meets natural language processing:
  {A} survey}.
\newblock {\it \bibinfo{journal}{CoRR}\/},  {\it
  \bibinfo{volume}{abs/2107.12603}\/}. \URLprefix
  \url{https://arxiv.org/abs/2107.12603}.
  \href{http://arxiv.org/abs/2107.12603}{\tt arXiv:2107.12603}.
\bibitem[{Mehnaz et~al.(2017)Mehnaz, Bellala \& Bertino}]{Meh17}
\bibinfo{author}{Mehnaz, S.}, \bibinfo{author}{Bellala, G.}, \&
  \bibinfo{author}{Bertino, E.} (\bibinfo{year}{2017}).
\newblock \bibinfo{title}{A secure sum protocol and its application to
  privacy-preserving multi-party analytics}.
\newblock In \bibinfo{editor}{E.~Bertino}, \bibinfo{editor}{R.~Sandhu}, \&
  \bibinfo{editor}{E.~R. Weippl} (Eds.), {\it \bibinfo{booktitle}{Proceedings
  of the 22nd {ACM} on Symposium on Access Control Models and Technologies,
  {SACMAT} 2017, Indianapolis, IN, USA, June 21-23, 2017}\/} (pp.
  \bibinfo{pages}{219--230}).
\newblock \bibinfo{publisher}{{ACM}}.
\newblock \DOIprefix\doi{10.1145/3078861.3078869}.
\bibitem[{Paillier(1999)}]{Pai99}
\bibinfo{author}{Paillier, P.} (\bibinfo{year}{1999}).
\newblock \bibinfo{title}{Public-key cryptosystems based on composite degree
  residuosity classes}.
\newblock In \bibinfo{editor}{J.~Stern} (Ed.), {\it
  \bibinfo{booktitle}{Advances in Cryptology - {EUROCRYPT} '99, Proceedings of
  International Conference on the Theory and Application of Cryptographic
  Techniques, Prague, Czech Republic, May 2-6}\/} (pp.
  \bibinfo{pages}{223--238}).
\newblock \bibinfo{publisher}{Springer} volume \bibinfo{volume}{1592} of {\it
  \bibinfo{series}{Lecture Notes in Computer Science}\/}.
\newblock \DOIprefix\doi{10.1007/3-540-48910-X\_16}.
\bibitem[{Phong et~al.(2018)Phong, Aono, Hayashi, Wang \& Moriai}]{Pho18}
\bibinfo{author}{Phong, L.~T.}, \bibinfo{author}{Aono, Y.},
  \bibinfo{author}{Hayashi, T.}, \bibinfo{author}{Wang, L.}, \&
  \bibinfo{author}{Moriai, S.} (\bibinfo{year}{2018}).
\newblock \bibinfo{title}{Privacy-preserving deep learning via additively
  homomorphic encryption}.
\newblock {\it \bibinfo{journal}{{IEEE} Trans. Information Forensics and
  Security}\/},  {\it \bibinfo{volume}{13}\/}, \bibinfo{pages}{1333--1345}.
  \DOIprefix\doi{10.1109/TIFS.2017.2787987}.
\bibitem[{Qi et~al.(2021)Qi, Zhou, Lei \& Zheng}]{Qi21}
\bibinfo{author}{Qi, J.}, \bibinfo{author}{Zhou, Q.}, \bibinfo{author}{Lei,
  L.}, \& \bibinfo{author}{Zheng, K.} (\bibinfo{year}{2021}).
\newblock \bibinfo{title}{Federated reinforcement learning: Techniques,
  applications, and open challenges}.
\newblock {\it \bibinfo{journal}{CoRR}\/},  {\it
  \bibinfo{volume}{abs/2108.11887}\/}. \URLprefix
  \url{https://arxiv.org/abs/2108.11887}.
  \href{http://arxiv.org/abs/2108.11887}{\tt arXiv:2108.11887}.
\bibitem[{Riazi et~al.(2019)Riazi, Samragh, Chen, Laine, Lauter \&
  Koushanfar}]{Ria19}
\bibinfo{author}{Riazi, M.~S.}, \bibinfo{author}{Samragh, M.},
  \bibinfo{author}{Chen, H.}, \bibinfo{author}{Laine, K.},
  \bibinfo{author}{Lauter, K.~E.}, \& \bibinfo{author}{Koushanfar, F.}
  (\bibinfo{year}{2019}).
\newblock \bibinfo{title}{{XONN:} xnor-based oblivious deep neural network
  inference}.
\newblock In \bibinfo{editor}{N.~Heninger}, \& \bibinfo{editor}{P.~Traynor}
  (Eds.), {\it \bibinfo{booktitle}{28th {USENIX} Security Symposium, {USENIX}
  Security 2019, Santa Clara, CA, USA, August 14-16, 2019}\/} (pp.
  \bibinfo{pages}{1501--1518}).
\newblock \bibinfo{publisher}{{USENIX} Association}.
\newblock \URLprefix
  \url{https://www.usenix.org/conference/usenixsecurity19/presentation/riazi}.
\bibitem[{Rumelhart et~al.(1986)Rumelhart, Hinton \& Williams}]{Rum86}
\bibinfo{author}{Rumelhart, D.~E.}, \bibinfo{author}{Hinton, G.~E.}, \&
  \bibinfo{author}{Williams, R.~J.} (\bibinfo{year}{1986}).
\newblock \bibinfo{title}{Learning representations by back-propagating errors}.
\newblock {\it \bibinfo{journal}{Nature}\/},  {\it \bibinfo{volume}{323}\/},
  \bibinfo{pages}{533--536}. \DOIprefix\doi{10.1038/323533a0}.
\bibitem[{Schlitter(2008)}]{Sch08}
\bibinfo{author}{Schlitter, N.} (\bibinfo{year}{2008}).
\newblock \bibinfo{title}{A protocol for privacy preserving neural network
  learning on horizontal partitioned data}.
\newblock \bibinfo{note}{Retrieved from
  \url{https://www.researchgate.net/publication/228944881_A_Protocol_for_Privacy_Preserving_Neural_Network_Learning_on_Horizontally_Partitioned_Data}}.
\bibitem[{Shokri \& Shmatikov(2015)}]{Sho15}
\bibinfo{author}{Shokri, R.}, \& \bibinfo{author}{Shmatikov, V.}
  (\bibinfo{year}{2015}).
\newblock \bibinfo{title}{Privacy-preserving deep learning}.
\newblock In \bibinfo{editor}{I.~Ray}, \bibinfo{editor}{N.~Li}, \&
  \bibinfo{editor}{C.~Kruegel} (Eds.), {\it \bibinfo{booktitle}{Proceedings of
  the 22nd {ACM} {SIGSAC} Conference on Computer and Communications Security,
  Denver, CO, USA, October 12-16, 2015}\/} (pp. \bibinfo{pages}{1310--1321}).
\newblock \bibinfo{publisher}{{ACM}}.
\newblock \DOIprefix\doi{10.1145/2810103.2813687}.
\bibitem[{Tang et~al.(2019)Tang, Wu, Liu, Wang \& Xian}]{Tan19}
\bibinfo{author}{Tang, F.}, \bibinfo{author}{Wu, W.}, \bibinfo{author}{Liu,
  J.}, \bibinfo{author}{Wang, H.}, \& \bibinfo{author}{Xian, M.}
  (\bibinfo{year}{2019}).
\newblock \bibinfo{title}{Privacy-preserving distributed deep learning via
  homomorphic re-encryption}.
\newblock {\it \bibinfo{journal}{Electronics}\/},  {\it \bibinfo{volume}{8}\/}.
  \DOIprefix\doi{10.3390/electronics8040411}.
\bibitem[{Urabe et~al.(2007)Urabe, Wang, Kodama \& Takata}]{Ura07}
\bibinfo{author}{Urabe, S.}, \bibinfo{author}{Wang, J.},
  \bibinfo{author}{Kodama, E.}, \& \bibinfo{author}{Takata, T.}
  (\bibinfo{year}{2007}).
\newblock \bibinfo{title}{A high collusion-resistant approach to distributed
  privacy-preserving data mining}.
\newblock In \bibinfo{editor}{H.~Burkhart} (Ed.), {\it
  \bibinfo{booktitle}{Proceedings of the {IASTED} International Conference on
  Parallel and Distributed Computing and Networks, as part of the 25th {IASTED}
  International Multi-Conference on Applied Informatics, February 13-15 2007,
  Innsbruck, Austria}\/} (pp. \bibinfo{pages}{307--312}).
\newblock \bibinfo{publisher}{{IASTED/ACTA} Press}.
\bibitem[{Vu et~al.(2020)Vu, Luong \& Ho}]{Vu20}
\bibinfo{author}{Vu, D.}, \bibinfo{author}{Luong, T.}, \& \bibinfo{author}{Ho,
  T.} (\bibinfo{year}{2020}).
\newblock \bibinfo{title}{An efficient approach for secure multi-party
  computation without authenticated channel}.
\newblock {\it \bibinfo{journal}{Information Sciences}\/},  {\it
  \bibinfo{volume}{527}\/}, \bibinfo{pages}{356--368}.
  \DOIprefix\doi{10.1016/j.ins.2019.07.031}.
\bibitem[{Wright \& Yang(2004)}]{Wri04}
\bibinfo{author}{Wright, R.~N.}, \& \bibinfo{author}{Yang, Z.}
  (\bibinfo{year}{2004}).
\newblock \bibinfo{title}{Privacy-preserving bayesian network structure
  computation on distributed heterogeneous data}.
\newblock In \bibinfo{editor}{W.~Kim}, \bibinfo{editor}{R.~Kohavi},
  \bibinfo{editor}{J.~Gehrke}, \& \bibinfo{editor}{W.~DuMouchel} (Eds.), {\it
  \bibinfo{booktitle}{Proceedings of the Tenth {ACM} {SIGKDD} International
  Conference on Knowledge Discovery and Data Mining, Seattle, Washington, USA,
  August 22-25, 2004}\/} (pp. \bibinfo{pages}{713--718}).
\newblock \bibinfo{publisher}{{ACM}}.
\newblock \DOIprefix\doi{10.1145/1014052.1014145}.
\bibitem[{Xu et~al.(2019)Xu, Baracaldo, Zhou, Anwar \& Ludwig}]{Xu19}
\bibinfo{author}{Xu, R.}, \bibinfo{author}{Baracaldo, N.},
  \bibinfo{author}{Zhou, Y.}, \bibinfo{author}{Anwar, A.}, \&
  \bibinfo{author}{Ludwig, H.} (\bibinfo{year}{2019}).
\newblock \bibinfo{title}{Hybridalpha: An efficient approach for
  privacy-preserving federated learning}.
\newblock In \bibinfo{editor}{L.~Cavallaro}, \bibinfo{editor}{J.~Kinder},
  \bibinfo{editor}{S.~Afroz}, \bibinfo{editor}{B.~Biggio},
  \bibinfo{editor}{N.~Carlini}, \bibinfo{editor}{Y.~Elovici}, \&
  \bibinfo{editor}{A.~Shabtai} (Eds.), {\it \bibinfo{booktitle}{Proceedings of
  the 12th {ACM} Workshop on Artificial Intelligence and Security, AISec@CCS
  2019, London, UK, November 15, 2019}\/} (pp. \bibinfo{pages}{13--23}).
\newblock \bibinfo{publisher}{{ACM}}.
\newblock \URLprefix \url{https://doi.org/10.1145/3338501.3357371}.
  \DOIprefix\doi{10.1145/3338501.3357371}.
\bibitem[{Yang et~al.(2019)Yang, Liu, Chen \& Tong}]{Yang20}
\bibinfo{author}{Yang, Q.}, \bibinfo{author}{Liu, Y.}, \bibinfo{author}{Chen,
  T.}, \& \bibinfo{author}{Tong, Y.} (\bibinfo{year}{2019}).
\newblock \bibinfo{title}{Federated machine learning: Concept and
  applications}.
\newblock {\it \bibinfo{journal}{{ACM} Trans. Intell. Syst. Technol.}\/},  {\it
  \bibinfo{volume}{10}\/}, \bibinfo{pages}{12:1--12:19}. \URLprefix
  \url{https://doi.org/10.1145/3298981}. \DOIprefix\doi{10.1145/3298981}.
\bibitem[{Yao(1982)}]{Yao82}
\bibinfo{author}{Yao, A.~C.} (\bibinfo{year}{1982}).
\newblock \bibinfo{title}{Protocols for secure computations (extended
  abstract)}.
\newblock In {\it \bibinfo{booktitle}{23rd Annual Symposium on Foundations of
  Computer Science, Chicago, Illinois, USA, 3-5 November 1982}\/} (pp.
  \bibinfo{pages}{160--164}).
\newblock \bibinfo{publisher}{{IEEE} Computer Society}.
\newblock \DOIprefix\doi{10.1109/SFCS.1982.38}.
\bibitem[{Zhu et~al.(2021)Zhu, Xu, Liu \& Jin}]{Zhu21}
\bibinfo{author}{Zhu, H.}, \bibinfo{author}{Xu, J.}, \bibinfo{author}{Liu, S.},
  \& \bibinfo{author}{Jin, Y.} (\bibinfo{year}{2021}).
\newblock \bibinfo{title}{Federated learning on non-iid data: {A} survey}.
\newblock {\it \bibinfo{journal}{CoRR}\/},  {\it
  \bibinfo{volume}{abs/2106.06843}\/}. \URLprefix
  \url{https://arxiv.org/abs/2106.06843}.
  \href{http://arxiv.org/abs/2106.06843}{\tt arXiv:2106.06843}.

\end{thebibliography}

\end{document}